\documentclass[aps,pre,a4paper,twocolumn,groupedaddress,showpacs,showkeys,preprintnumbers,floatfix,dvips,superscriptaddress]{revtex4-1}

\usepackage[utf8]{inputenc} \usepackage[T1]{fontenc}
\usepackage{psfrag} 
\usepackage{times} \usepackage{graphicx} \usepackage{color}
\usepackage{amsmath} \usepackage{amscd} \usepackage{amssymb} \usepackage{amsthm}
\usepackage{ragged2e} \usepackage{subfigure} \usepackage{caption}
\usepackage{nicefrac}
\usepackage{captcont}
\usepackage{verbatim}

\DeclareCaptionJustification{reallyjustified}{\justifying}
\begin{document}

\title{An analytical expression for the exit probability of the $q$-voter model in one dimension}
\author{Andr\'e M. Timpanaro}
\email[]{timpa@if.usp.br}
\affiliation{Instituto de F\'{i}sica, Universidade de S\~{a}o Paulo Caixa Postal 66318, 05314-970 - S\~{a}o Paulo - S\~{a}o Paulo - Brazil}
\affiliation{CEVIPOF - SciencesPo and CNRS -  Paris, 98 rue de l'Universit\'e, 75007 - France}
\author{Serge Galam}
\email[]{serge.galam@sciencespo.fr}
\affiliation{CEVIPOF - SciencesPo and CNRS -  Paris, 98 rue de l'Universit\'e, 75007 - France}
\date{\today}


\begin{abstract}
We present in this paper an approximation that is able to give an analytical expression for the exit probability of the $q$-voter model in one dimension. This expression gives a better fit for the more recent data about simulations in large networks \cite{q-voter-Timpanaro}, and as such, departs from the expression $\frac{\rho^q}{\rho^q + (1-\rho)^q}$ found in papers that investigated small networks only \cite{q-voter-Sznajd, q-voter-Slanina, q-voter-Lambiotte}. The approximation consists in assuming a large separation on the time scales at which active groups of agents convince inactive ones and the time taken in the competition between active groups. Some interesting findings are that for $q=2$ we still have $\frac{\rho^2}{\rho^2 + (1-\rho)^2}$ as the exit probability and for large values of $q$ the difference between the result and $\frac{\rho^q}{\rho^q + (1-\rho)^q}$ becomes negligible (the difference is maximum for $q=5$ and 6)
\end{abstract}

\maketitle

\section{Introduction}

In the last years, the study of sociophysics has applied tools from statistical physics to the study of social phenomena, leading to some insights on the origins of some of the phenomena studied by sociologists and political scientists \cite{galam-book}. At the same time, by taking statistical physics far from its usual domain of application new challenges arise that are by themselves interesting to study as they could reveal unknown aspects of ithe theory that could be used again in physical systems. This work concerns one of those challenges, the controversy around the exit probability of the one dimensional $q$-voter model.

The $q$-voter model is an opinion propagation model defined in \cite{q-votante-def}, where groups of $q$ agreeing agents are needed for opinion propagation to occur. A series of papers \cite{q-voter-Sznajd, q-voter-Slanina, q-voter-Lambiotte, q-voter-Galam} studied this model in one dimension and a controversy about its exit probability (the probability that a given opinion becomes the dominant one as a function of its starting proportion of agents in an uncorrelated initial condition) sparked. In a recent paper \cite{q-voter-Timpanaro}, one of the authors made simulations of the model in large networks, showing that the expression fitted in \cite{q-voter-Sznajd}, $\frac{\rho^q}{\rho^q + (1-\rho)^q}$ is a very good approximation, but deviations were found for $q = 3,4$ and 5 (but not for $q=2$). Some justification was given for this expression when $q$ was large and a Kirkwood approximation yields the same result for $q=2$, however no general deduction for the expression was given. Also, since this expression is not completely accurate (as found from the simulations), a treatment able to find those corrections is desirable.

On this paper, we build on the basic idea of the duel model (defined in \cite{q-voter-Timpanaro}) that is used for the large network simulations, to get an  approximation for the exit probability for an uncorrelated initial condition, that can be calculated anaytically in the thermodynamical limit. We compare this expression with the simulation results obtained in \cite{q-voter-Timpanaro} and show that it gives a much better fit of the data.

\subsection{Model Definition}

The $q$-voter model as studied in this paper is defined in a linear chain, where each site is an agent that has an opinion that can be either + or -. The time evolution is given as follows:

\begin{itemize}
\item At each time step, choose a site $i$ and $q$ of its neighbours (consecutively), that is; $i-1, i-2, \ldots, i-q$ or $i+1, i+2, \ldots, i+q$.
\item If the neighbours have all the same opinion, $i$ copies their opinion. Otherwise nothing happens.
\end{itemize}

As was shown in \cite{q-voter-Timpanaro}, the model can be equally described in terms of contiguous groups of agreeing agents (the dual model). Here a group of size $n$ and spin $s$ means a sequence of $n$ neighbouring sites (that can't be made larger) with all of them having spin $s$ (for example, this is what a group of size 3 and spin + means: $\ldots - [+ + +] - \ldots$). The  rules for the dual description are:

\begin{itemize}
\item Choose a group $i$ such that it's size $n_i$ is at least $q$.
\item Choose $s=\pm 1$.
\item Pass one agent from $i+s$ to $i$, that is $n_i\rightarrow n_i +1$ and $n_{i+s} \rightarrow n_{i+s} - 1$.
\item If this causes $n_{i+s} = 0$, Remove group $i+s$ and merge groups $i$ and $i+2s$ (adding their sizes together).
\end{itemize}

\section{The Approximation}
\label{sec:approximation}

The dual formulation of the model makes it clear that the dynamics hapens on the borders of the active groups (that is, groups with at least $q$ agents). However there is a difference in the interaction between two active groups and the interaction between an active and an inactive group.

When two active groups "compete", the border between them undergoes an unbiased random walk, while when one of the groups is inactive the border always moves "invading" the inactive group. This means that the time needed for an active group to destroy an inactive group is much smaller than the time needed to destroy an active one (or even to make a comparable change on the size of an active group).

We make then the approximation that while there are any inactive groups, the borders between active groups remain static, and all the borders between an active and an inactive group move at the same speed.

This approximation makes the first part of the transient, where the agents coarse-grain into active groups, purely deterministic.

The second part of the transient is very similar to the voter model, with the difference being what happens when a group drops below $q$ agents. According to our approximation, the $q-1$ remaining agents would be absorbed, so to keep this part also deterministic (instead of dependant on the order in which the groups are destroyed) we will simply neglect these left over agents (that is, they are removed as soon as their group becomes inactive), which is the same as removing $q-1$ sites from each group after the first part of the transient is over and then following the usual voter model ($q=1$).

Since the voter model has a trivial exit probability, the exit probability can be calculated directly from the initial condition, which can be done analytically in the thermodynamic limit. We do this calculation in the following section, but we also provide an algorithm implementing the approximation in finite chains (using the same ideas presented in the next section) in appendix \ref{ap:code}.

\section{Deduction of the analytical expression}
\label{ap:deducao}

To make the calculation of the exit probability according to our approximation we must find how the transformation discussed in section \ref{sec:approximation} behaves in an uncorrelated initial condition. This can be done by making the transformation on the fly as the initial condition is generated and keeping track of the number of $+$ and $-$ sites after the transformation. To do this we consider the spin patterns that can occur as the initial condition is drawn. We are going to denote by $(m,n,s_1,s_2)$ the pattern where the last active group drawn had spin $s_1$, followed by an inactive region containing $m$ sites with spin $+$ and $n$ with spin $-$, and followed by an active group with spin $s_2$. These patterns are transitions between a sequence with $q$ equal sites to the next sequence containing $q$ equal sites, so we will denote a group with more than $q$ sites using the patterns $(0,0,+,+)$ and $(0,0,-,-)$ (for example a group with $q+5$ sites having spin $-$ is denoted by 5 consecutive $(0,0,-,-)$ patterns).

When $q$ is bigger than 2, most of the patterns represent more than one way the initial condition can be drawn. For example, if $q=3$, both $+++--+-++---$ and $+++-+-+-+---$ are patterns of type $(3,3,+,-)$. Each of these particular ways a pattern can be drawn are equiprobable and they all give the same end result when the transformation is applied (For $(m,n,+,-)$ and $(m,n,-,+)$ patterns this is a consequence of each step in the expansion of the relevant active groups conserving the number of $+$ and $-$ sites), so all that we need to keep track is their multiplicity and the probability weight of each of them.

The probability weights are straightforward. With the exception of $(0,0,+,+)$ and $(0,0,-,-)$ we have
\[
P(m,n,s_1,s_2) = \rho^m (1-\rho)^n P(s_1) P(s_2),
\]
where $P(+) = \rho$ and $P(-) = 1-\rho$. For $(0,0,+,+)$ and $(0,0,-,-)$ the probabilities are $\rho^{q+1}$ and $(1-\rho)^{q+1}$ respectively. The multiplicities $\Omega(m,n,s_1,s_2)$ must obey the following recurrence relation (the patterns $(0,0,s,s)$ must be treated differently, but its trivial that they always have multiplicity 1)

\begin{equation}
\Omega(m,n,s_1,s_2) = \sum_{r=1}^{q-1} \sum_{t=1}^{q-1} \Omega(m-r,n-t,s_1,s_2),
\label{eq:recurrence}
\end{equation}
the reasoning being that the inactive part of a pattern $(m,n,s_1,s_2)$ can be formed by drawing $r$ sites  with spin $-s_1$, followed by $t$ sites with spin $s_1$, followed by any way that the inactive part of a pattern $(m-r,n-t,s_1,s_2)$ can be drawn, so that we must add the multiplicities of all possibilities.

The patterns $(0,0,s,s)$ don't fit in this reasoning and because of this for the purpose of the recurrence relation we must take $\Omega(0,0,+,+) = \Omega(0,0,-,-) = 0$. Obviously, we must also take $\Omega(m,n,s,t) = 0$ whenever $m$ or $n$ is negative. The rest of the initial conditions are

\[
\left\{
\begin{array}{ll}
\Omega(0,0,+,-) = \Omega(0,0,-,+) = 1 &\\
\Omega(0,m,+,-) = \Omega(m,0,-,+) = 0 &\mbox{ if }m\neq 0\\
\Omega(0,m,+,+) = \Omega(m,0,-,-) = 1 &\mbox{ if }1\leq m\leq q-1\\
\Omega(0,m,+,+) = \Omega(m,0,-,-) = 0 &\mbox{ if }m\geq q\\
\end{array}
\right.
\]

Equation \ref{eq:recurrence} cannot be solved analytically, however only the generating functions 
\begin{equation}
\varphi_{s_1,s_2}(x,y) = \sum_{m,n} \Omega(m,n,s_1,s_2) x^m y^n
\label{eq:generating-def}
\end{equation}
turn out to be relevant. It is easy to show that these are

\begin{equation}
\varphi_{+,+}(x,y) = \frac{\Phi(x)}{1 - \Phi(x)\Phi(y)}
\label{eq:generating++}
\end{equation}
\begin{equation}
\varphi_{+,-}(x,y) = \varphi_{-,+}(x,y) = \frac{1}{1 - \Phi(x)\Phi(y)}
\label{eq:generating+-}
\end{equation}
\begin{equation}
\varphi_{-,-}(x,y) = \frac{\Phi(y)}{1 - \Phi(x)\Phi(y)}
\label{eq:generating--}
\end{equation}
where

\[
\Phi(x) = \frac{x^q - 1}{x - 1}
\]

Finally, we need the increase $(\Delta_+, \Delta_-)$ that a pattern will be responsible for. These are

\[
\left\{
\begin{array}{ll}
\Delta(m,n,+,+) = (m+n+q,0) & \mbox{ if }m,n\neq 0 \\
\Delta(m,n,-,-) = (0,m+n+q) & \mbox{ if }m,n\neq 0 \\
\Delta(0,0,+,+) = (1,0)&\\
\Delta(0,0,-,-) = (0,1)&\\
\Delta(m,n,+,-) = (m,n+1)&\\
\Delta(m,n,-,+) = (m+1,n)&
\end{array}
\right.
\]
Note that in the last 2 cases we are preemptively taking into account the $q-1$ sites that get removed from each group after the expansion of the active groups finishes.

Putting it all together, we have the following proportions after accounting for all processes

\[
(N_+(\rho), N_-(\rho)) = (\rho^{q+1}, (1-\rho)^{q+1}) + 
\]\[
 + \sum_{m,n = 0}^{\infty}\sum_{s_1,s_2} P(m,n,s_1,s_2)\Omega(m,n,s_1,s_2)\Delta(m,n,s_1,s_2),
\]
where the $\Omega$ denote the solutions of equation \ref{eq:recurrence}. By the symmetry of the problem we must have $N_+(\rho) = N_-(1-\rho)$. Consider then the function

\[
K(x,y) = x^{q+1} + x^{2q}\sum_{m,n} \Omega(m,n,+,+)x^m y^n(m+n+q) + 
\]\[
+ 2x^qy^q \sum_{m,n} \Omega(m,n,+,-)x^m y^n(2m+1),
\]
we must have then $N_+(\rho) = K(\rho, 1-\rho)$. However, $K$ can be rewritten as

\[
K(x,y) = x^{q+1} + (q + x\partial_x + y\partial_y) \varphi_{+,+}(x,y) + 
\]
\begin{equation}
 +(1 + 2x\partial_x) \varphi_{+,-}(x,y).
\label{eq:K}
\end{equation}

Inputing then equations \ref{eq:generating++} and \ref{eq:generating+-} into equation \ref{eq:K} and normalizing

\[
E(\rho) = \frac{K(\rho)}{K(\rho) + K(1-\rho)}
\]
gives the exit probability. By performing the algebraic manipulations, one arrives at the result

\begin{widetext}
\[
E(\rho)=\frac{\rho^q - \rho^q(1-(1-\rho)^{q-1})^2 + \rho^{2q-2} - \rho^{2q-2}(1-\rho)^{q-1}(2\rho^2 - \rho + 1 + 2q\rho(1-\rho)) + \rho^{2q-1}(1-\rho)^{2q-2}(\rho+q(1-\rho))}{(\rho^{q-1} + (1-\rho)^{q-1} - \rho^{q-1}(1-\rho)^{q-1})(\rho^{q-1} + (1-\rho)^{q-1} - \rho^{q-1}(1-\rho)^{q-1} - 2(q-1)\rho^q(1-\rho)^q)}
\]
\end{widetext}

\section{Comparison with simulation results}

The exit probability that we found in the last section is

\[
E(\rho) = \frac{K(\rho)}{K(\rho) + K(1-\rho)}, 
\]
where

\[
K(\rho) = \rho^q - \rho^q(1-(1-\rho)^{q-1})^2 + \rho^{2q-2} +
\]\[
 - \rho^{2q-2}(1-\rho)^{q-1}(2\rho^2 - \rho + 1 + 2q\rho(1-\rho)) + 
\]
\begin{equation}
 +\rho^{2q-1}(1-\rho)^{2q-2}(\rho+q(1-\rho))
\end{equation}

It's interesting to see that for $q>2$ we have
\[
K(\rho) = \rho^q + \mathcal{O}(\rho^{q+1}),
\]
which explains why
\[
E(\rho) = \frac{\rho^q}{\rho^q + (1-\rho)^q}
\]
is such a good approximation. Moreover, for $q=2$:

\[
K(\rho) = \rho^2 + 2\rho^2(1-\rho)-\rho^2(1-\rho)^2 + \rho^2(-2\rho^3+5\rho^2-\rho-1)+ 
\]\[
 +\rho^3(1-\rho)^2(2-\rho) = \rho^2(1-\rho^2(1-\rho)^2) \Rightarrow 
\]\[ 
\Rightarrow K(1-\rho) = (1-\rho)^2(1-\rho^2(1-\rho)^2) 
\]
and hence
\[
E(\rho) = \frac{\rho^2}{\rho^2 + (1-\rho)^2}.
\]

Also, if $q=1$
\[
K(\rho) = \rho + 1 + \frac{\rho^2-1}{1-\rho} +\rho =
\]\[
= 2\rho + 1 - (\rho + 1) = \rho
\]
implying $E(\rho) = \rho$ as expected.

\begin{figure}[hbt!]
\includegraphics[width=\columnwidth]{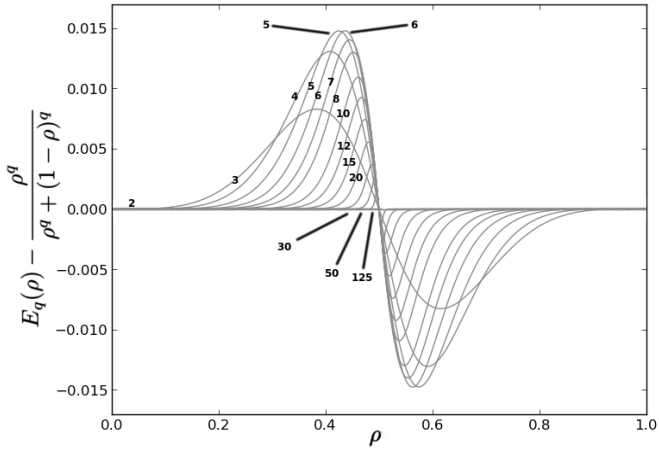}
\caption{Differences between the exit probability predicted by our approximation and the formula proposed in \cite{q-voter-Sznajd}. Note that as $q$ becomes larger the 2 formulas become increasingly closer, while being exactly the same for $q=2$ and having noticeable differences for intermediate values of $q$. The values of $q$ plotted are 2, 3, 4, 5, 6, 7, 8, 10, 12, 15, 20, 30, 50 and 125.}
\label{fig:curves}
\end{figure}

On figure \ref{fig:curves} we have the curves $E(\rho) - \frac{\rho^q}{\rho^q + (1-\rho)^q}$ to show how the difference behaves as $q$ increases, showing that $E(\rho)$ approaches $\frac{\rho^q}{\rho^q + (1-\rho)^q}$ once again for large $q$.

\begin{figure}[hbt!]
\includegraphics[width=\columnwidth]{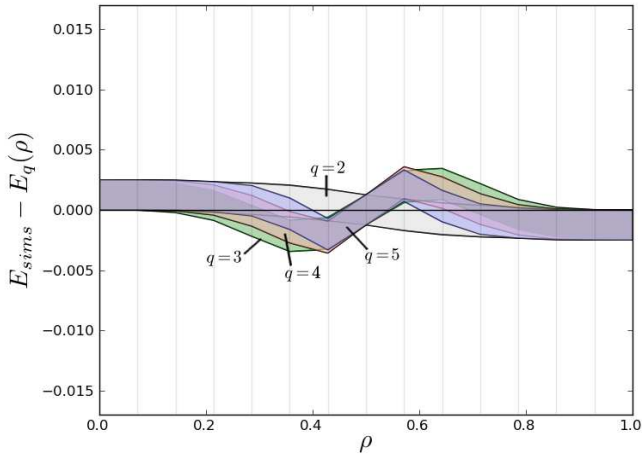}
\caption{(Colour online) Difference between the simulation results and the exit probability predicted from our approximation. The bands represent the region where the procedure done in \cite{q-voter-Timpanaro} estimates the true difference to be (statistical errors are too small to be seen in the graph). The network size being compared is $3.16\times 10^7$ sites for the values $q=2,3,4$ and 5 (gray, green, red and blue respectively). Note that for $q\neq 2$ there is still a small discrepancy around $\rho = 0.4$ and $\rho = 0.6$.}
\label{fig:data}
\end{figure}

On figure \ref{fig:data} we have a comparison between our expression and the data obtained in \cite{q-voter-Timpanaro}

\section{Conclusion}

We have presented an approximation that is able to explain most of the discrepancies found between the exit probability in simulations done in \cite{q-voter-Timpanaro} and the expression proposed in \cite{q-voter-Sznajd}. The only things that the approximation assumes is that there is a complete separation on the time scales between active-active and active-inactive group interactions, and that the sites that are left over when a group ceases to be active can be neglected

{\bf Acknowledgements}

André Martin Timpanaro would like to acknowledge FAPESP for financial support.

\appendix

\section{Python script implementing the approximation}
\label{ap:code}

\noindent\verb;import random, sys;\\
\\
\verb;#Argument usage:;\\
\verb;#<n (int)> <q (int)> <rho (float)>;\\
\\
\verb;n       = int(sys.argv[1]);\\
\verb;q       = int(sys.argv[2]);\\
\verb;rho     = float(sys.argv[3]);\\
\\
\verb;opposite  = (1, 0);\\
\verb;M = {((0,0),1,1):0, ((0,0),0,0):0};\\
\verb;#the keys are ((n+, n-), sl, sr);\\
\\
\verb;def rand_spin():;\\
\textvisiblespace\verb;rand = random.uniform(0, 1);\\
\textvisiblespace\verb;if rand < rho:;\\
\textvisiblespace\textvisiblespace\verb;return 1  #+;\\
\textvisiblespace\verb;else:;\\
\textvisiblespace\textvisiblespace\verb;return 0  #-;\\
\\
\\
\verb;last    = 1;\\
\verb;#the spin of the last active group;\\
\verb;curr    = 1;\\
\verb;#the spin of the current group;\\
\verb;spin    = 0;\\
\verb;#the spin that was drawn;\\
\verb;size    = 0;\\
\verb;#size of the group;\\
\verb;i       = 0;\\
\verb;aux     = [0, 0];\\
\verb;while i < n:   #measure M;\\
\textvisiblespace\verb;if spin != curr:;\\
\textvisiblespace\textvisiblespace\verb;if size >= q:;\\
\textvisiblespace\textvisiblespace\textvisiblespace\verb;if last == curr: #++ or --;\\
\textvisiblespace\textvisiblespace\textvisiblespace\textvisiblespace\verb;group = tuple(aux);\\
\textvisiblespace\textvisiblespace\textvisiblespace\textvisiblespace\verb;if (group, last, last) in M:;\\
\textvisiblespace\textvisiblespace\textvisiblespace\textvisiblespace\textvisiblespace\verb;M[(group, last, last)] += 1;\\
\textvisiblespace\textvisiblespace\textvisiblespace\textvisiblespace\verb;else:;\\
\textvisiblespace\textvisiblespace\textvisiblespace\textvisiblespace\textvisiblespace\verb;M[(group, last, last)] = 1;\\
\textvisiblespace\textvisiblespace\textvisiblespace\textvisiblespace\verb;M[((0,0), last, last)] += size-q;\\
\textvisiblespace\textvisiblespace\textvisiblespace\verb;else: #+- or -+;\\
\textvisiblespace\textvisiblespace\textvisiblespace\textvisiblespace\verb;group = tuple(aux);\\
\textvisiblespace\textvisiblespace\textvisiblespace\textvisiblespace\verb;if (group, last, curr) in M:;\\
\textvisiblespace\textvisiblespace\textvisiblespace\textvisiblespace\textvisiblespace\verb;M[(group, last, curr)] += 1;\\
\textvisiblespace\textvisiblespace\textvisiblespace\textvisiblespace\verb;else:;\\
\textvisiblespace\textvisiblespace\textvisiblespace\textvisiblespace\textvisiblespace\verb;M[(group, last, curr)] = 1;\\
\textvisiblespace\textvisiblespace\textvisiblespace\textvisiblespace\verb;M[((0,0), curr, curr)] += size-q;\\
\textvisiblespace\textvisiblespace\textvisiblespace\textvisiblespace\verb;last = curr;\\
\textvisiblespace\textvisiblespace\textvisiblespace\verb;aux  = [0, 0];\\
\textvisiblespace\textvisiblespace\verb;else:;\\
\textvisiblespace\textvisiblespace\textvisiblespace\verb;aux[curr] += size;\\
\textvisiblespace\textvisiblespace\verb;curr = spin;\\
\textvisiblespace\textvisiblespace\verb;size = 0;\\
\textvisiblespace\verb;size += 1;\\
\textvisiblespace\verb;spin = rand_spin();\\
\textvisiblespace\verb;i += 1;\\
\verb;N = [0, 0];\\
\verb;for ((m,n),sl,sr) in M:;\\
\textvisiblespace\verb;occurences = M[((m,n),sl,sr)];\\
\textvisiblespace\verb;if (m,n) == (0,0):;\\
\textvisiblespace\textvisiblespace\verb;N[sl] += occurences;\\
\textvisiblespace\verb;elif sl != sr:;\\
\textvisiblespace\textvisiblespace\verb;N[0]  += m*occurences;\\
\textvisiblespace\textvisiblespace\verb;N[1]  += n*occurences;\\
\textvisiblespace\textvisiblespace\verb;N[sr] += occurences;\\
\textvisiblespace\verb;else:;\\
\textvisiblespace\textvisiblespace\verb;N[sr] += (m+n+q)*occurences;\\
\verb;E = float(N[1])/float(N[0] + N[1]);\\
\verb;print E;\\

\bibliographystyle{plain}
\bibliography{andre}

\end{document}